\documentclass[10pt,conference]{IEEEtran}
\usepackage{graphicx} 
\usepackage{indentfirst}
\usepackage{algpseudocode}
\usepackage{amsmath}
\usepackage{amsfonts}
\usepackage{amssymb}
\usepackage{multirow}
\usepackage{algorithm}
\usepackage[font=small]{caption}
\usepackage{cite}
\usepackage{pifont}
\usepackage{makecell}
\usepackage{subcaption}
\usepackage{enumitem}
\usepackage{float}
\usepackage{url}
\def\BibTeX{{\rm B\kern-.05em{\sc i\kern-.025em b}\kern-.08em T\kern-.1667em\lower.7ex\hbox{E}\kern-.125emX}}

\title{Generalized User-Oriented Image Semantic Coding Empowered by Large Vision-Language Model}
\author{
    \IEEEauthorblockN{Sin-Yu Huang and Vincent W.S. Wong}
        Department of Electrical and Computer Engineering, The University of British Columbia, Vancouver, Canada\\
        Email: \{syhuang, vincentw\}@ece.ubc.ca
}

\begin{document}

\maketitle

\begin{abstract}
Semantic communication has shown outstanding performance in preserving the overall source information in wireless transmission. For semantically rich content such as images, human users are often interested in specific regions
depending on their intent. Moreover, recent semantic coding models are mostly trained on specific datasets. However, real-world applications may involve images out of the distribution of training dataset, which makes generalization a crucial but largely unexplored problem. To incorporate user’s intent into semantic coding, in this paper, we propose a generalized user-oriented image semantic coding (UO-ISC) framework, where the user provides a text query indicating its intent. The transmitter extracts features from the source image which are relevant to the user’s query. The receiver reconstructs an image based on those features. To enhance the generalization ability, we integrate contrastive language image pre-training (CLIP) model, which is a pretrained large vision-language model (VLM), into our proposed UO-ISC framework. To evaluate the relevance between the reconstructed image and the user’s query, we introduce the user-intent relevance loss, which is computed by using a pretrained large VLM, large language-and-vision assistant (LLaVA) model. When performing zero-shot inference on unseen objects, simulation results show that the proposed UO-ISC framework outperforms the state-of-the-art query-aware image semantic coding in terms of the answer match rate.
\end{abstract}

\section{Introduction}
\label{Sec:intro}

The advent of intelligent applications, such as autonomous vehicles, augmented reality/virtual reality (AR/VR), and remote surgery, has driven the demand for higher data rates, lower latency, and higher reliability. To meet these requirements, the sixth-generation (6G) wireless networks are anticipated to shift from the traditional data-oriented communication to semantic communication~\cite{Getu2024}. In semantic communication, the goal is to transmit information that preserves the meaning and relevance of the intended task~\cite{Deniz2023,Zhang2024SCAN}. 
In~\cite{Huang2023}, Huang \textit{et al.} introduced the rate-semantic-perceptual loss, which jointly evaluates the bit rate, semantic loss, and task-execution loss. However, the tasks considered in~\cite{Huang2023}, such as detection, segmentation and classification, are all machine-centric and do not fully reflect human-centric communication scenarios, where the intent is often complex and dynamic. For example, given an image of a horse on a grassy ranch with people nearby, a rancher may ask: ``Where is my horse?'', focusing on the animal. On the other hand, the parents may ask: ``Are my kids in the farm?'', prioritizing the people. This example illustrates the importance of semantic content depends on the user's intent. Thus, semantic coding should adaptively prioritize the content relevant to the user's intent. This capability is crucial for applications such as AR/VR and surveillance, where user's intent may change depending on the scenarios.

Recently, foundation models, such as ChatGPT and Gemini, have shown outstanding reasoning capabilities. Some studies (e.g.,~\cite{Jiang2025,Zhao2024}) have leveraged foundation models in semantic coding. Other related works have integrated foundation models into semantic coding guided by the user’s query. In~\cite{chen2024}, Chen \textit{et al.} used a large language model (LLM) to generate answers and select key semantic elements from images based on user's query. However, they did not consider the generalization ability on unseen objects. In real-world applications, the input data are diverse and may include previously unseen objects, making generalization a critical challenge for semantic coding. In our previous work~\cite{Huang2025}, we explored the zero-shot capability of the semantic coding for text transmission. However, the generalization ability of image-based semantic coding remains largely unexplored.
\begin{figure*}
    \centering
    \includegraphics[width=0.98\linewidth]{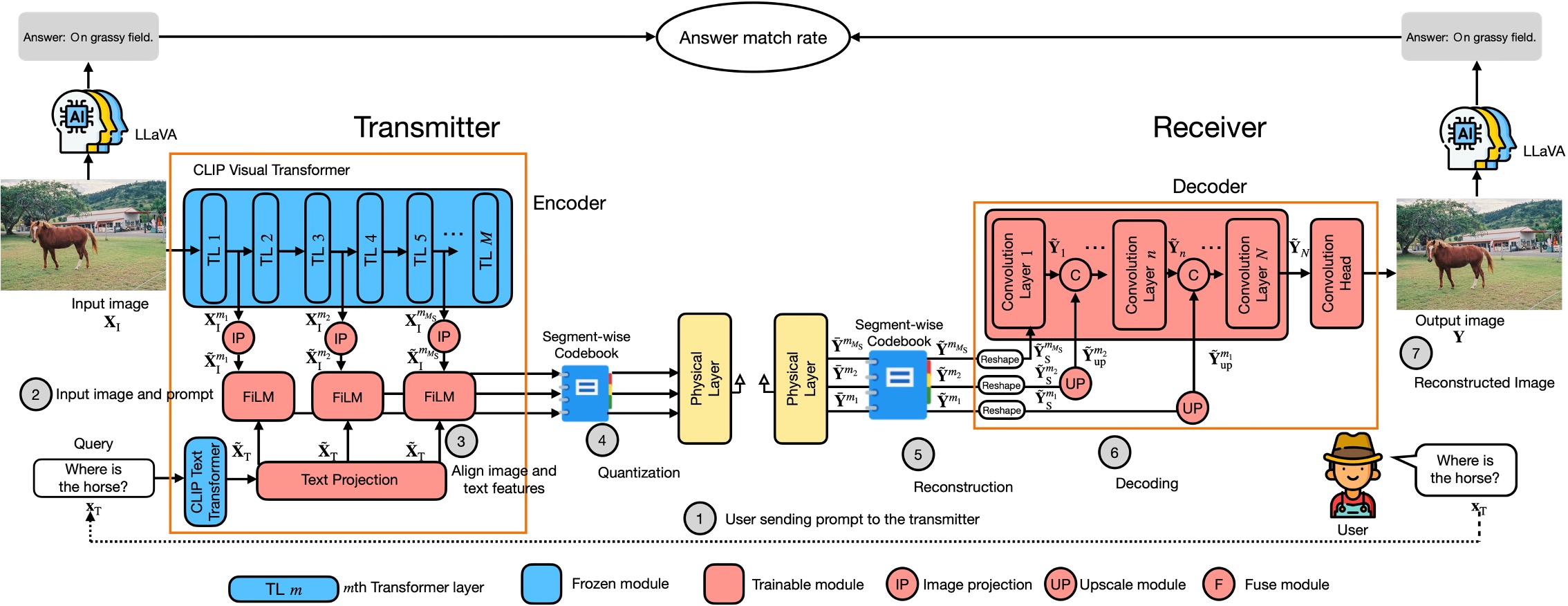}
    \caption{The system model of the proposed generalized UO-ISC framework. As an example, we consider $\mathcal{M}_{\text{S}}=\{1,3,5\}$ in the transmitter.}
    \vspace{-15pt}
    \label{fig:system_model}
\end{figure*}

In this paper, we aim to answer the following questions:
\begin{enumerate}[label=Q\arabic*:]
    \item \textit{When transmitting an image with semantically rich content, how to incorporate user's intent into semantic coding?}
    \item \textit{How to improve the generalization capability of semantic coding?}
\end{enumerate}
The main contributions of this paper are as follows:
\begin{itemize}
    \item We propose a generalized user-oriented image semantic coding (UO-ISC) framework, which prioritizes the semantic features relevant to user's intent for wireless transmission. To incorporate the user's intent, the transmitter takes the image and text query as input. To improve the generalization capability, we use contrastive language-image pre-training (CLIP)~\cite{Alec2021}, which is a pretrained large vision-language model (VLM), to extract the features from the image and text query. These features are aligned, partitioned into segments, and quantized using a segment-wise codebook. The receiver of the UO-ISC framework reconstructs the image based on the received semantic features. 
    
    \item We propose a user-intent relevance loss function to evaluate the relevance between the generated image and the original image for a given text query. The loss is determined by large language-and-vision assistant (LLaVA) model. Moreover, we propose a two-phase optimization algorithm to jointly train the model using the user-intent relevance loss, $\ell_1$ loss, quantization loss, and adversarial training~\cite{Good2014}, aiming to enhance both the semantic relevance with the user's query and the image quality through discrete feature representations.

    \item For performance evaluation, we train our proposed UO-ISC framework on the visual question answering (VQA) dataset~\cite{Stan2015} and perform zero-shot evaluation. When comparing the answers between the reconstructed and original images with the same query, results show that integrating CLIP improves the answer match rate by 34\%. Furthermore, incorporating user intent provides an additional 5\% improvement over the model without text query alignment and a 4.8\% improvement over the state-of-the-art query-aware image semantic coding~\cite{chen2024}.
\end{itemize}

This paper is organized as follows. Section~\ref{Sec:system} presents the system model of the proposed UO-ISC framework. Section~\ref{Sec:problem} introduces the user-intent relevance loss and two-phase training algorithm for UO-ISC. The performance evaluation is given in Section~\ref{Sec:simulation}. Section~\ref{Sec:conclude} concludes the paper.

\textit{Notations:} We use boldface lowercase letters to denote vectors, boldface uppercase letters to denote matrices, and calligraphic letters to represent sets. For a matrix $\mathbf{X}$, we denote its transpose as $\mathbf{X}^{T}$, its $i$th row as vector $\mathbf{x}_i$, and the $(i,j)$th element as $[\mathbf{X}]_{i,j}$. The expectation operator is denoted by $\mathbb{E}[\cdot]$, and $\odot$ indicates element-wise multiplication. The $\ell_1$-norm and $\ell_2$-norm are denoted by $\|\cdot\|_1$ and $\|\cdot\|_2$, respectively. The dot product of vectors $\mathbf{a}$ and $\mathbf{b}$ is denoted by $\langle \mathbf{a}, \mathbf{b} \rangle$.

\section{System Model}
\label{Sec:system}
Our proposed generalized UO-ISC framework is illustrated in Fig.~\ref{fig:system_model}. It consists of a transmitter and a receiver. The transmitter obtains an image from sensing or stored files. Before transmission, the user provides a text query indicating the intended semantic content of the image to the transmitter based on prior knowledge of the scene. The transmitter then takes both the image and the text query as input. The goal is to reconstruct an image at the receiver while taking into account the semantic content relevant to the user's intent. 

Specifically, in the transmitter, we leverage CLIP~\cite{Alec2021}, which is a large VLM pretrained on extensive image-text pairs, to extract the features from the text query and the input image. Let $\mathbf{x}_{\text{T}}$ denote the input text query. Let $\mathbf{X}_{\text{I}} \in \mathbb{R}^{C \times H \times W}$ denote the input image, where $C$, $H$, and $W$ are the number of channels, the dimensions along the height and width directions, respectively. The text query is processed by the CLIP text transformer to extract the text feature matrix $\mathbf{X}_{\text{T}} \in \mathbb{R}^{N_{\text{T}} \times D}$, where $N_{\text{T}}$ is the number of text embeddings and $D$ is the embedding dimension. The image is processed by the CLIP visual transformer, which consists of $M$ transformer layers. For each layer $m \in \mathcal{M} = \{1, 2, \ldots, M\}$, the output feature matrix is denoted by $\mathbf{X}_{\text{I}}^{m} \in \mathbb{R}^{N_{\text{I}} \times D}$, where $N_{\text{I}}$ is the number of image embeddings. Inspired by skip connections in U-Net~\cite{Ronne2015}, we forward the features from different encoder layers to the decoder in order to preserve the spatial details for better reconstruction. However, transmitting the features from all transformer layers incurs a large transmission overhead. Thus, we only select a subset of transformer layers indexed by $\mathcal{M}_{\text{S}}=\{m_1,\dots,m_{M_{\text{S}}}\} \subseteq \mathcal{M}$, where $M_{\text{S}}=|\mathcal{M}_{\text{S}}| \leq M $. Only the feature matrices $\mathbf{X}_{\text{I}}^{m}$, for $m\in \mathcal{M}_{\text{S}}$, are processed in the subsequent transmission pipeline. Note that all transformer layers in the CLIP visual and text encoders are kept frozen to preserve the pretrained model's generalization ability.

To extract the image features which are relevant to the user's intent, we align the text and image features by projecting both modalities into a shared embedding space using a multilayer perceptron (MLP). The projected image and text feature matrices are given by:
\begin{align}
    \mathbf{\tilde{X}}_{\text{I}}^{m} &= \mathcal{P}_{\text{I}}^{m}(\mathbf{X}_{\text{I}}^{m}; \boldsymbol{\theta}^{m}_{\text{P, I}}) \in \mathbb{R}^{N_{\text{I}} \times D}, \quad m\in \mathcal{M}_{\text{S}}, \\
    \mathbf{\tilde{X}}_{\text{T}} &= \mathcal{P}_{\text{T}}(\mathbf{X}_{\text{T}}; \boldsymbol{\theta}_{\text{P, T}}) \in \mathbb{R}^{N_{\text{T}} \times D}, 
\end{align}
where $\mathcal{P}_{\text{I}}^{m}(\cdot;\, \boldsymbol{\theta}^{m}_{\text{P, I}})$ and $\mathcal{P}_{\text{T}}(\cdot;\, \boldsymbol{\theta}_{\text{P, T}})$ are the projection functions for the $m$th selected image features and text features with parameter sets $\boldsymbol{\theta}^{m}_{\text{P, I}}$ and $\boldsymbol{\theta}_{\text{P, T}}$, respectively. Each projected image feature matrix $\mathbf{\tilde{X}}_{\text{I}}^{m}$ is paired with the projected text feature matrix $\mathbf{\tilde{X}}_{\text{T}}$. They are processed by Feature-wise Linear Modulation (FiLM)~\cite{Dum2018} to align the two modalities. Specifically, $\mathbf{\tilde{X}}_{\text{T}}$ is projected through two MLPs to generate the scale matrix $\mathbf{\tilde{X}}_{\text{T,sc}}^{m}\in \mathbb{R}^{N_{\text{I}} \times D}$ and shift matrix $\mathbf{\tilde{X}}_{\text{T,sh}}^{m} \in \mathbb{R}^{N_{\text{I}} \times D}$: 
\begin{align}
    \mathbf{\tilde{X}}_{\text{T,sc}}^{m} = \mathcal{E}_{\text{sc}}^{m}(\mathbf{\tilde{X}}_{\text{T}};\boldsymbol{\theta}_{\text{E,sc}}^{m}),~ 
    \mathbf{\tilde{X}}_{\text{T,sh}}^{m} = \mathcal{E}_{\text{sh}}^{m}(\mathbf{\tilde{X}}_{\text{T}};\boldsymbol{\theta}_{\text{E,sh}}^{m}),
    \end{align}
where \(\mathcal{E}_{\text{sc}}^{m}(\cdot;\boldsymbol{\theta}_{\text{E,sc}}^{m})\) and \(\mathcal{E}_{\text{sh}}^{m}(\cdot;\boldsymbol{\theta}_{\text{E,sh}}^{m})\) denote the MLP with parameters \(\boldsymbol{\theta}_{\text{E,sc}}^{m}\) and \(\boldsymbol{\theta}_{\text{E,sh}}^{m}\), respectively. The aligned feature matrix $\mathbf{\tilde{X}}_{\text{align}}^{m}$ is given by:
\begin{align} 
    \mathbf{\tilde{X}}_{\text{align}}^{m} = \mathbf{\tilde{X}}_{\text{T,sc}}^{m} \odot \mathbf{\tilde{X}}_{\text{I}}^{m}+\mathbf{\tilde{X}}_{\text{T,sh}}^{m}  \in \mathbb{R}^{N_{\text{I}} \times D}, \quad m\in \mathcal{M}_{\text{S}}. \label{eq:align}
\end{align} 
\begin{figure}[t]
    \centering
    \includegraphics[width=0.8\linewidth]{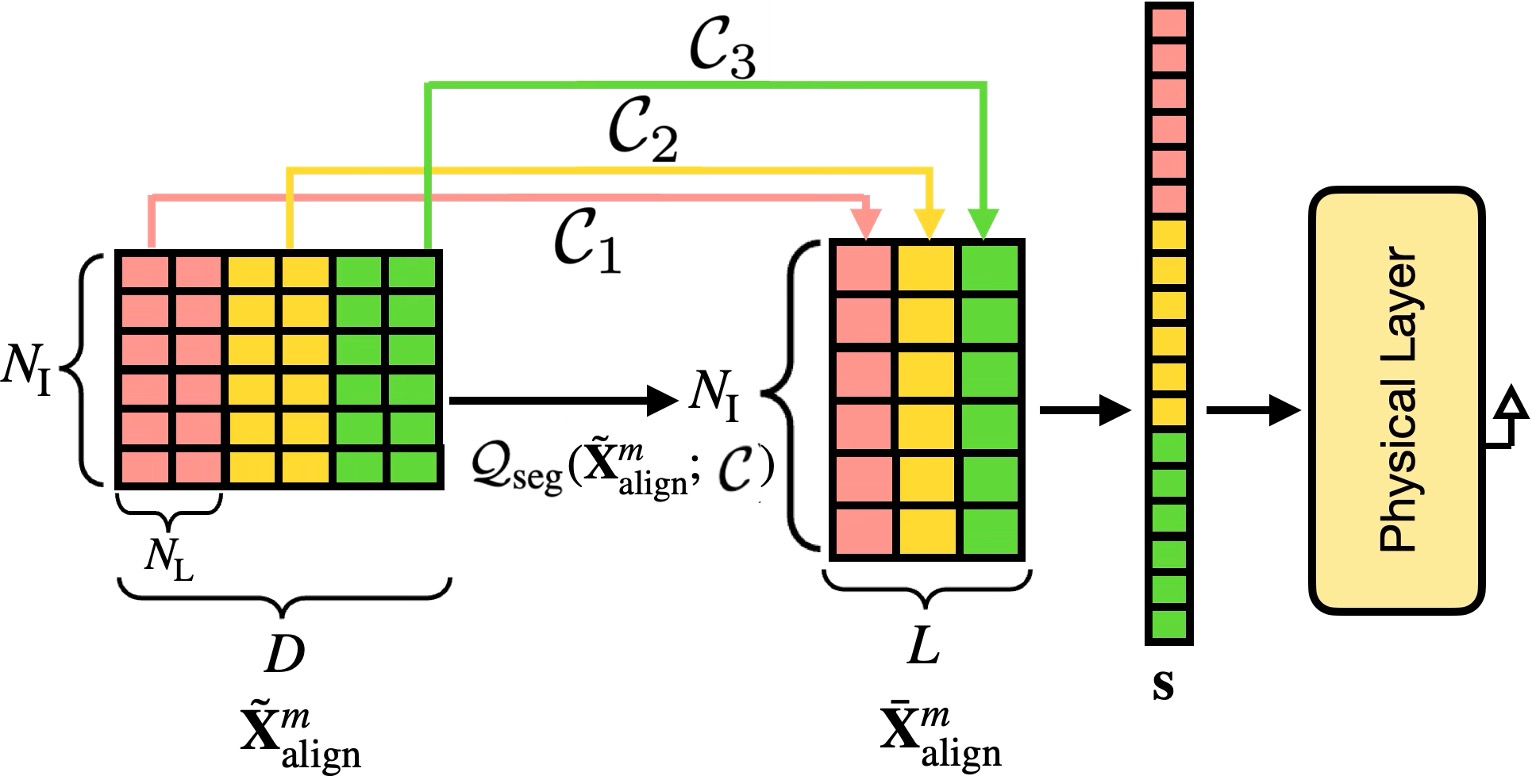}
    \caption{The segment-wise vector quantization process in UO-ISC.}
    \label{fig:transfer}
    \vspace{-15pt}
\end{figure}
To further compress the features, we apply vector quantization to convert each embedding vector in matrix $\mathbf{\tilde{X}}_{\text{align}}^{m}$ into a sequence of discrete integers. We partition each embedding vector into multiple segments and quantize each segment independently using its own set of codewords. As illustrated in Fig.~\ref{fig:transfer}, for an embedding vector with dimension $D$, we partition it into $L$ segments, where each segment has length $N_{\text{L}}=\frac{D}{L}$. We assume $D$ is divisible by $L$, so that the embedding vector can be partitioned into segments of equal length. The $l$th codebook $\mathcal{C}_{l} = \left\{ \mathbf{c}_i^l \in \mathbb{R}^{1 \times N_{\text{L}}} \right\}_{i=1}^{N_{\text{cw}}}$ consists of a set of codewords $\mathbf{c}_i^l$, where $N_{\text{cw}}$ is the number of codewords. For the aligned feature matrix $\mathbf{\tilde{X}}_{\text{align}}^{m} \in \mathbb{R}^{N_{\text{I}} \times D}$, we obtain the quantized feature matrix $\mathbf{\bar{X}}_{\text{align}}^{m}$ by segment-wise vector quantization function $\mathcal{Q}_{\text{seg}}(\cdot;\mathcal{C})$, where $\mathcal{C} = \{ \mathbf{c}_i^{l} \, |\, i = 1, \ldots, N_{\text{cw}},\ l = 1, \ldots, L \}$ is the set of all codebooks. We have
\begin{align}
    \mathbf{\bar{X}}_{\text{align}}^{m} = \mathcal{Q}_{\text{seg}}(\mathbf{\tilde{X}}_{\text{align}}^{m}; \mathcal{C}) \in \{1, \ldots, N_{\text{cw}}\}^{N_{\text{I}} \times L}, m\in \mathcal{M}_{\text{S}}. \label{eq:quant}
\end{align}
The segment-wise vector quantization partitions each row feature vector $\mathbf{\tilde{x}}_{\text{align}}^{m} \in \mathbb{R}^{1 \times D}$ in the aligned feature  matrix $\mathbf{\tilde{X}}_{\text{align}}^{m}$ into $L$ segments of length $N_{\text{L}}$, and assigns each segment to its nearest codeword in the codebook $\mathcal{C}_{l}$. The operation of function  $\mathcal{Q}_{\text{seg}}(\mathbf{\tilde{X}}_{\text{align}}^{m}; \mathcal{C})$ is as follows. For each row index $n \in \{1, \ldots, N_{\text{I}}\}$ and segment index $l \in \{1, \ldots, L\}$, we extract the $l$th segment from the $n$th row vector of $\mathbf{\tilde{X}}_{\text{align}}^{m}$ as
\begin{align}
    \mathbf{u}_{n,l}^{m} = \big( 
        &[\mathbf{\tilde{X}}_{\text{align}}^{m}]_{n,\,(l-1)L + 1},\ 
         [\mathbf{\tilde{X}}_{\text{align}}^{m}]_{n,\,(l-1)L + 2},\ \notag \\
        &\dots,\ 
         [\mathbf{\tilde{X}}_{\text{align}}^{m}]_{n,\,(l-1)L + N_{\text{L}}} 
    \big) \in \mathbb{R}^{1 \times N_{\text{L}}}. \label{eq:segment}
\end{align}
We obtain the quantized matrix $\mathbf{\bar{X}}_{\text{align}}^{m}$, where each quantized index is given by
\begin{align}
    \left[ \mathbf{\bar{X}}_{\text{align}}^{m} \right]_{n,l} &= 
    \arg\min_{i \in \{1, \ldots, N_{\text{cw}}\}} 
    \left\| \mathbf{u}_{n,l}^{m} - \mathbf{c}_i^l \right\|_2^2, \notag \\
    &\quad n \in \{1, \ldots, N_{\text{I}}\},\;
    l \in \left\{1, \ldots,L \right\}. \label{eq:quant_index}
\end{align}

After quantization, the quantized feature matrices $\{\mathbf{\bar{X}}_{\text{align}}^{m}\}_{m \in \mathcal{M}_{\text{S}}}$ are then flattened and concatenated into a single symbol vector $\mathbf{s} \in \{1, \ldots, N_{\text{cw}}\}^{M_{\text{S}} N_{\text{I}} L}$. This vector is then passed through the physical layer, including channel coding and modulation. The resulting complex-valued transmit signal is denoted by $\mathbf{s}_{\text{x}} \in \mathbb{C}^{N_{\text{S}}}$, where $N_{\mathrm{S}}$ is the number of transmitted symbols after physical layer processing. Consider a flat fading channel model, the received signal is given by
\begin{align}
    \mathbf{\hat{s}}_{\text{x}} = h \mathbf{s}_{\text{x}} + \mathbf{n}, \label{eq:channel}
\end{align}
where $h$ is the complex channel gain, and $\mathbf{n} \sim \mathcal{CN}(\mathbf{0}, \sigma^{2} \mathbf{I}_{N_{\mathrm{S}}})$ denotes the additive white Gaussian noise with noise variance $\sigma^2$.

At the receiver side, the received signal $\mathbf{\hat{s}}_{\mathrm{x}}$ is processed in the physical layer module, including demodulation and channel decoding, to determine the discrete symbol vector $\mathbf{\hat{s}} \in \{1, \ldots, N_{\text{cw}}\}^{M_{\text{S}} N_{\text{I}} L}$. The vector $\hat{\mathbf{s}}$ is then reshaped back into a collection of quantized feature matrices $\{\mathbf{\bar{Y}}_{\text{align}}^{m}\}_{m \in \mathcal{M}_{\text{S}}}$, where each matrix $\mathbf{\bar{Y}}_{\text{align}}^{m} \in \{1, \ldots, N_{\text{cw}}\}^{N_{\text{I}} \times L}$.
Then, the segment-wise reconstruction function $\mathcal{Q}_{\text{seg}}^{-1}(\cdot;\, \mathcal{C})$ is applied to each quantized feature matrix $\mathbf{\bar{Y}}_{\text{align}}^{m}$. This function reverses the quantization process by mapping each quantized element into its corresponding codeword in the codebook. Let the element in the $n$th row and $l$th column of $\mathbf{\bar{Y}}_{\text{align}}^{m}$ be denoted by $\bar{y}_{n,l}^{m} \in \{1, \ldots, N_{\text{cw}}\}$. The reconstructed embedding matrix, denoted by $\mathbf{\tilde{Y}}_{\text{align}}^{m} = \mathcal{Q}^{-1}(\bar{\mathbf{Y}}_{\text{align}}^{m};\, \mathcal{C})$, is given by
\begin{align}
    \mathbf{\tilde{Y}}_{\text{align}}^{m} =
    \begin{bmatrix}
        \mathbf{c}^{1}_{\bar{y}_{1,1}^{m}} & \mathbf{c}^{2}_{\bar{y}_{1,2}^{m}} & \cdots & \mathbf{c}^{L}_{\bar{y}_{1,L}^{m}} \\
        \mathbf{c}^{1}_{\bar{y}_{2,1}^{m}} & \mathbf{c}^{2}_{\bar{y}_{2,2}^{m}} & \cdots & \mathbf{c}^{L}_{\bar{y}_{2,L}^{m}} \\
        \vdots & \vdots & \ddots & \vdots \\
        \mathbf{c}^{1}_{\bar{y}_{N_{\text{I}},1}^{m}} & \mathbf{c}^{2}_{\bar{y}_{N_{\text{I}},2}^{m}} & \cdots & \mathbf{c}^{L}_{\bar{y}_{N_{\text{I}},L}^{m}}
    \end{bmatrix}
    \in \mathbb{R}^{N_{\text{I}} \times D},
    \label{eq:dequant_matrix}
\end{align}
where $\mathbf{c}^{l}_{\bar{y}_{n,l}^{m}} \in \mathbb{R}^{1 \times N_{\text{L}}}$ is the codeword associated with index $\bar{y}_{n,l}^{m}$ from the $l$th codebook $\mathcal{C}_{l}= \{ \mathbf{c}_i^l \in \mathbb{R}^{1 \times N_{\text{L}}} \}_{i = 1}^{N_{\text{cw}}}$. 

After obtaining the reconstructed embeddings $\mathbf{\tilde{Y}}_{\text{align}}^{m}$ for all $m \in \mathcal{M}_{\text{S}}$, each matrix is reshaped into a spatial feature tensor $\mathbf{\tilde{Y}}_{\text{S}}^{m} \in \mathbb{R}^{\tilde{H} \times \tilde{W} \times D}$, where $\tilde{H}\tilde{W}=N_{\text{I}}$. These spatial feature tensors are then passed through a decoder composed of $N$ convolution layers, indexed by $n \in \mathcal{N}=\{1, \ldots, N\}$. Each convolutional layer is represented by function $\mathcal{D}_{n}(\cdot;\, \boldsymbol{\phi}_{\text{D}}^{n})$, parameterized by $\boldsymbol{\phi}_{\text{D}}^{n}$, and produces an output $\mathbf{\tilde{Y}}^{n}\in \mathbb{R}^{\tilde{H}_{n} \times \tilde{W}_{n} \times D}$, where the spatial dimensions $\tilde{H}_{n}$ and $\tilde{W}_{n}$ are progressively increase for each layer. Each convolutional block consists of two standard convolution layers and a transposed convolution to perform spatial upsampling, where each standard convolution layer is followed by instance normalization and a leaky rectified linear unit (ReLU) activation. The number of output channels is equal to $D$ for all layers. The decoder processes the embeddings in reverse order, starting from the reconstructed feature tensor of the final transformer layer's output $\mathbf{\tilde{Y}}_{\text{S}}^{m_{M_{\text{S}}}}$. For $n = 1$, the feature tensor $\mathbf{\tilde{Y}}_{\text{S}}^{m_{M_{\text{S}}}}$ is passed directly to the first convolutional layer. \begin{samepage}For $n>1$, the output from the previous layer $\mathbf{\tilde{Y}}^{n-1}\vspace{6pt}$ is fused with the upsampled version of the corresponding reconstructed embedding, defined as $ \mathbf{\tilde{Y}}_{\text{up}}^{m_{M_{\text{S}} - n + 1}} = \mathcal{U}_{n}( \mathbf{\tilde{Y}}_{\text{S}}^{m_{M_{\text{S}} - n + 1}}; \boldsymbol{\phi}_{\text{U}}^{n} )$, where $\mathcal{U}_{n}(\cdot; \boldsymbol{\phi}_{\text{U}}^{n})$ denotes a learnable upsampling module.\end{samepage} This upsampled tensor is then fused with $\mathbf{\tilde{Y}}^{n-1}$ via a feature fusion MLP, defined as $\mathbf{\tilde{Y}}_{\text{fused}}^{n-1} = \mathcal{F}_{n}\left( \mathbf{\tilde{Y}}_{\text{up}}^{m_{M_{\text{S}} - n + 1}}, \mathbf{\tilde{Y}}^{n-1};\boldsymbol{\phi}_{\text{F}}^{n} \right)$ where $\mathcal{F}_{n}(\cdot, \cdot;, \boldsymbol{\phi}_{\text{F}}^{n})$ merges the aligned features and intermediate output. The tensor $\mathbf{\tilde{Y}}_{\text{fused}}^{n-1}$ is then passed to the $n$th convolutional layer
\begin{align}
    \mathbf{\tilde{Y}}^{n} = 
    \begin{cases}
        \mathcal{D}_{n}\left( \mathbf{\tilde{Y}}_{\text{S}}^{m_{M_{\text{S}}}};\, \boldsymbol{\phi}_{\text{D}}^{n} \right), & n = 1, \\[1ex]
        \mathcal{D}_{n}\left( \mathbf{\tilde{Y}}_{\text{fused}}^{n-1};\, \boldsymbol{\phi}_{\text{D}}^{n} \right), & n \in \mathcal{N} \setminus \{1\}.
    \end{cases}
    \label{eq:decoder}
\end{align}
In our work, the number of convolution layers $N$ is equal to the number of selected indices $M_{\text{S}}$ so that each transmitted feature can combine with the convolution output. The final decoded embeddings $\mathbf{\tilde{Y}}^{N}$ pass through the final convolution head, generating the reconstructed image $\mathbf{Y}$. Note that all the trainable parameters in the encoder is included in the parameter set $\boldsymbol{\theta} = \{ \boldsymbol{\theta}^{m}_{\text{P, I}}, \boldsymbol{\theta}_{\text{P, T}}, \boldsymbol{\theta}^{m}_{\text{E,sc}}, \boldsymbol{\theta}^{m}_{\text{E,sh}} \}_{m\in\mathcal{M}_{\text{S}}}$ and all the trainable parameters in the decoder is included in the parameter set $\boldsymbol{\phi} = \{ \boldsymbol{\phi}^{n}_{\text{U}}, \boldsymbol{\phi}^{n}_{\text{F}}, \boldsymbol{\phi}^{n}_{\text{D}}\}_{n\in\mathcal{N}}$.

\section{Problem Formulation and the Proposed Two-phase Training Algorithm} 
\label{Sec:problem} 
The proposed UO-ISC framework reconstructs the source image while prioritizing the semantic features relevant to the user's intent, where the intent is specified by a text query. To evaluate the alignment between the reconstructed image with the user's intent, we define the \textit{user-intent relevance loss}, which is determined using the pretrained LLaVA model~\cite{Liu2023}. LLaVA is a large VLM that integrates a pretrained visual encoder and an LLM via a lightweight projection module. To assess semantic alignment, both the original image $\mathbf{X}_{\text{I}}$ and the reconstructed image $\mathbf{Y}$ are paired with the same text query $\mathbf{x}_{\text{T}}$ and processed separately through LLaVA. The images are processed through the visual encoder $\mathcal{F}_{\text{v}}(\cdot)$ and projection matrix $\mathbf{P}_{\text{v}}$ to obtain $\mathbf{H}_{\text{X}} = \mathbf{P}_{\text{v}} \mathcal{F}_{\text{v}}(\mathbf{X}_{\text{I}})$ and $\mathbf{H}_{\text{Y}} = \mathbf{P}_{\text{v}} \mathcal{F}_{\text{v}}(\mathbf{Y})$.
The text query is tokenized into $\mathbf{H}_{\text{T}}$, and concatenated with the image embeddings to form the multimodal sequences $\mathbf{H}_{\text{X}}^{\text{M}}=[\mathbf{H}_{\text{X}}, \mathbf{H}_{\text{T}}]$ and $\mathbf{H}_{\text{Y}}^{\text{M}}=[\mathbf{H}_{\text{Y}}, \mathbf{H}_{\text{T}}]$, which are input to the language model. The output hidden states are denoted by $\mathbf{\tilde{H}}_{\text{X}}^{\text{M}} = [\mathbf{\tilde{H}}_{\text{X}}, \mathbf{\tilde{H}}_{\text{T}}] $ and $\mathbf{\tilde{H}}_{\text{Y}}^{\text{M}} = [\mathbf{\tilde{H}}_{\text{Y}}, \mathbf{\tilde{H}}_{\text{T}}] $. The user-intent relevance loss is defined as the cosine distance between the corresponding image tokens:
\begin{align}
    \mathcal{L}_{\text{user}}(\mathbf{Y}, \mathbf{X}_{\text{I}};\, \mathbf{x}_{\text{T}}) 
    = 1 - \frac{1}{N_{\text{i}}} \sum_{i=1}^{N_{\text{i}}} \frac{
        \langle\mathbf{\tilde{h}}_{\text{X}, i}^{\text{M}}, \mathbf{\tilde{h}}_{\text{Y}, i}^{\text{M}}\rangle
    }{
        \| \mathbf{\tilde{h}}_{\text{X}, i}^{\text{M}} \|_2 
        \| \mathbf{\tilde{h}}_{\text{Y}, i}^{\text{M}} \|_2
    },
\end{align}
where $\mathbf{\tilde{h}}_{\text{X}, i}^{\text{M}}$ and $\mathbf{\tilde{h}}_{\text{Y}, i}^{\text{M}}$ denote the $i$th token embedding vector from $\mathbf{\tilde{H}}_{\text{X}}^{\text{M}}$ and $\mathbf{\tilde{H}}_{\text{Y}}^{\text{M}}$, respectively, and $N_{\text{i}}$ denotes the total number of token embedding vectors.

In addition to user-intent relevance, the visual quality of the reconstructed image is also important for image semantic coding. To improve the visual quality, we incorporate the $\ell_1$ loss and adversarial loss~\cite{Good2014} into the training objective of UO-ISC. The $\ell_1$ loss, defined as $\mathcal{L}_{\text{1}}(\mathbf{X}_{\text{I}}, \mathbf{Y})= \|\mathbf{X}_{\text{I}} - \mathbf{Y}\|_{1}$, improves the pixel-level fidelity between the source and reconstructed images, and preserves low-level visual details such as edges and structural content. The adversarial loss improves the perceptual realism of the reconstructed images. In adversarial training, a lightweight discriminator is used to distinguish real images from the generated ones. The generator, which is our model, is trained to produce outputs that are indistinguishable from real images. Additionally, the segment-wise codebook $\mathcal{C}$ introduces a quantization loss. Let $q \in \{1,\ldots, N_{\text{cw}}\}$ denote the index of the closest codeword obtained from~\eqref{eq:quant_index} and $\mathbf{c}_q^{l} \in \mathcal{C}_l$ denote the nearest codeword in the $l$th segment. Recall from~\eqref{eq:segment} that $\mathbf{u}_{n,l}^{m}$ is the $l$th segment of the $n$th row vector in $\mathbf{\tilde{X}}_{\text{align}}^{m}$. The quantization loss is defined as
\begin{align}
    \mathcal{L}_{\text{quant}} 
    &= \sum_{m \in \mathcal{M}_{\text{S}}} \sum_{n=1}^{N_{\text{I}}} \sum_{l=1}^{L} \Big( 
    \left\| \text{sg}[\mathbf{u}_{n,l}^{m}] - \mathbf{c}_q^{l} \right\|_2^2 \notag \\
    &\hspace{5em}
    + \beta
    \left\| \mathbf{u}_{n,l}^{m} - \text{sg}[\mathbf{c}_q^{l}] \right\|_2^2 
    \Big),
    \label{eq:L_quant}
\end{align}
where $\text{sg}[\cdot]$ denotes the stop-gradient operator and $\beta$ is a tunable parameter. Since the codeword selection function in~\eqref{eq:quant} is non-differentiable, the stop-gradient operator ensures separate updates to the encoder and codebook during backpropagation.
\begin{figure*}[t]
    \centering
    \includegraphics[width=\linewidth]{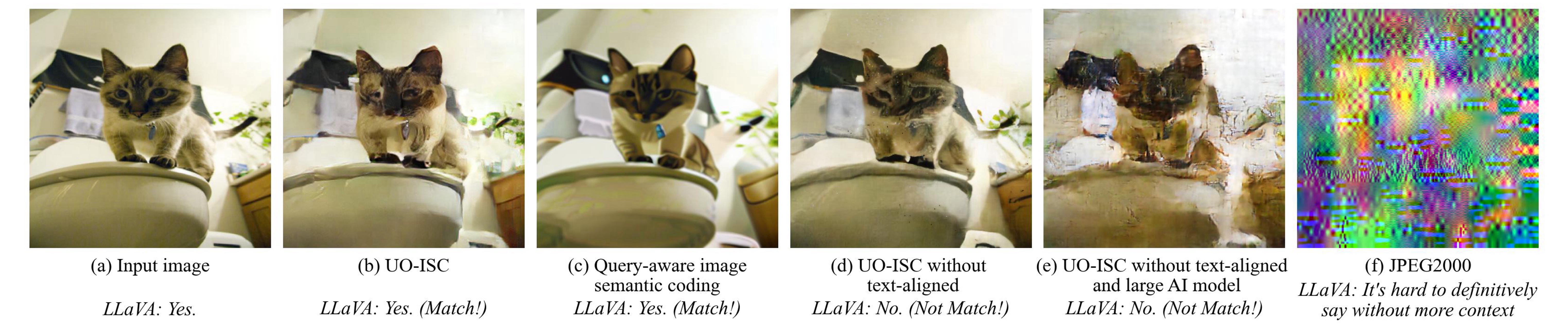}
    \caption{\small Visualization of the reconstructed images produced by the proposed UO-ISC and baseline methods with the SNR = 0 dB. The text query used for evaluation is: ``Does the cat have a collar?''.}
    \vspace{-15pt}
    \label{fig:visual}
\end{figure*}

Since jointly optimizing all loss terms may lead to unstable training, we adopt a two-phase training strategy. In Phase I, the adversarial loss is not considered, and the model is optimized using the following loss function:
\begin{align}
    \mathcal{L}_{\text{phase-I}}(\mathbf{X}_{\text{I}}, \mathbf{Y};\, \mathbf{x}_{\text{T}}) 
    =\ & \mathcal{L}_{1}(\mathbf{X}_{\text{I}}, \mathbf{Y}) 
    + \lambda_{\text{user}}\, \mathcal{L}_{\text{user}}(\mathbf{X}_{\text{I}}, \mathbf{Y};\, \mathbf{x}_{\text{T}}) \notag \\
    & + \lambda_{\text{quant}}\, \mathcal{L}_{\text{quant}}, \label{eq:loss_phase1}
\end{align}
where $\lambda_{\text{user}}$ and $\lambda_{\text{quant}}$ are tunable parameters. This phase enables the model to learn semantic reconstruction from discrete features, even though the resulting images may lack visual sharpness. In Phase II, we introduce adversarial training to enhance perceptual realism. A lightweight discriminator $\mathcal{D}_{\text{dis}}(\cdot\,;\boldsymbol{\psi}_{\text{dis}})$ with parameter set $\boldsymbol{\psi}_{\text{dis}}$ is trained to distinguish real images from the generated images. The adversarial loss function for the generator is defined as $\mathcal{L}_{\text{gen}}(\mathbf{Y})=\mathbb{E}[\log\mathcal{D}_{\text{dis}}(\mathbf{Y};\boldsymbol{\psi}_{\text{dis}})] $, which enables the generator to produce images to be indistinguishable from real samples. The overall loss function for Phase II is as follows: 
\begin{align}
    \mathcal{L}_{\text{phase-II}}(\mathbf{X}_{\text{I}}, \mathbf{Y};\, \mathbf{x}_{\text{T}}) 
    &= \mathcal{L}_{\text{phase-I}}(\mathbf{X}_{\text{I}}, \mathbf{Y};\, \mathbf{x}_{\text{T}}) \nonumber \\
    &\quad + \lambda_{\text{gen}}\, \mathcal{L}_{\text{gen}}(\mathbf{Y}), \label{eq:loss_phase2}
\end{align}
where $\lambda_{\text{gen}}$ is a tunable parameter. After updating the generator, the discriminator is trained using the following loss function:
\begin{align}
    \mathcal{L}_{\text{dis}}(\mathbf{X}_{\text{I}},\mathbf{Y}) 
    &= -\mathbb{E}[\log \mathcal{D}_{\text{dis}}(\mathbf{X}_{\text{I}};\boldsymbol{\psi}_{\text{dis}})] \nonumber \\
    &\quad - \mathbb{E}[\log(1 - \mathcal{D}_{\text{dis}}(\mathbf{Y};\boldsymbol{\psi}_{\text{dis}}))], \label{Eq:discriminatorloss}
\end{align}
where the first term encourages high confidence for real images $\mathbf{X}_{\text{I}}$, and the second term penalizes for assigning high confidence to the reconstructed output $\mathbf{Y}$.
In Phase II, the training process alternates between our model and the discriminator. First, we update our model and keep the discriminator to be fixed. Then, we update the discriminator while keeping our model to be fixed. The complete training procedure is summarized in Algorithm~\ref{alg:stage1}.

\begin{algorithm}[t]
\small
\caption{\small Proposed Two-phase Training Algorithm}\label{alg:stage1}
\begin{algorithmic}[1]
\State \textbf{Input:} Image data and query in the training datasets; number of epochs for the first and second phases, $N_1$ and $N_2$; learning rate for our model and the discriminator $\eta_{\text{gen}}$ and $\eta_{\text{dis}}$; regularization weights $\lambda_{\text{user}}$, $\lambda_{\text{gen}}$, $\lambda_{\text{quant}}$ and $\beta$.
\State \textbf{Phase I:}
\State Freeze the parameters of pretrained CLIP in the encoder.
\For{$i \gets 1$ \textbf{to} $N_1$} 
    \State Load input image $\mathbf{X}_{\text{I}}$ and query $\mathbf{x}_{\text{T}}$ from the dataset.
    \State Sample the channel gain coefficient $h$ and noise $\mathbf{n}$.
    \State Perform forward propagation.
    \State Calculate the loss $\mathcal{L}_{\text{phase-I}}(\mathbf{X}_{\text{I}}, \mathbf{Y};\, \mathbf{x}_{\text{T}})$ based on~\eqref{eq:loss_phase1}.
    \State Update parameters $\{\boldsymbol{\theta}, \boldsymbol{\phi}\}$ and codewords in $\mathcal{C}$ \text{by }$\mathcal{L}_{\text{phase-I}}$.
\EndFor
\State \textbf{Phase II:}
\State Load parameter sets $\{\boldsymbol{\theta}, \boldsymbol{\phi}, \mathcal{C}\}$ trained in the first phase.
\For{$i \gets 1$ \textbf{to} $N_2$} 
    \State Load input image $\mathbf{X}_{\text{I}}$ and query $\mathbf{x}_{\text{T}}$ from the dataset.
    \State Sample the channel gain coefficient $h$ and noise $\mathbf{n}$.
    \State Perform forward propagation.
    \State Calculate the loss $\mathcal{L}_{\text{phase-II}}(\mathbf{X}_{\text{I}}, \mathbf{Y};\, \mathbf{x}_{\text{T}})$ based on~\eqref{eq:loss_phase2}.
    \State Update parameters $\{\boldsymbol{\theta}, \boldsymbol{\phi}\}$ and codewords in $\mathcal{C}$ \text{by }$\mathcal{L}_{\text{phase-II}}$.
    \State Perform forward propagation on discriminator.
    \State Calculate the loss $\mathcal{L}_{\text{disc}}(\mathbf{X}_{\text{I}},\mathbf{Y})$ based on (\ref{Eq:discriminatorloss}).
    \State Update the discriminator parameters $\boldsymbol{\psi}_{\text{dis}}$ \text{by} $\mathcal{L}_{\text{disc}}$.

\EndFor
\State \textbf{Output}: The optimized parameters $\{\boldsymbol{\theta}^{*}, \boldsymbol{\phi}^{*}\}$ and codebook $\mathcal{C} = \{ {\mathbf{c}_i^{l}}^{*} \, |\, i = 1, \ldots, N_{\text{cw}},\ l = 1, \ldots, L \}$.
\end{algorithmic}
\end{algorithm}

\section{Performance Evaluation}
\label{Sec:simulation}
In this section, we evaluate the performance of the proposed UO-ISC framework and compare it with four baseline models. The UO-ISC framework employs the pretrained CLIP ViT-B/16\footnote{Model available at https://github.com/openai/CLIP} model with a patch size of 16 and a projection dimension of $D = 512$. In the encoder, we select the transformer layers $\mathcal{M}_{\text{S}} = \{3, 6, 9, 11\}$ from a total of $M = 12$ layers, resulting in $N = 4$ convolutional layers in the decoder. Each feature vector is partitioned into segments of length $N_{\text{L}} = 32$, leading to $L = \frac{D}{N_{\text{L}}} = 16$ segments per vector. Each segment is quantized using a codebook containing $N_{\text{cw}} = 64$ codewords. We use low density parity check (LDPC) codes and 8-ary modulation. We consider a flat fading channel with $h \sim \mathcal{CN}(0,1)$. We utilize nanoLLaVA\footnote{Model available at \url{https://huggingface.co/qnguyen3/nanoLLaVA}} to determine the user-intent relevance loss $\mathcal{L}_{\text{user}}$. The UO-ISC framework is trained on the VQA dataset~\cite{Stan2015}. We compare its performance with the following baseline models:
\begin{itemize}
    \item \textbf{UO-ISC without text alignment:} This model adopts the same architecture as UO-ISC but does not include the feature alignment with text query and does not consider the user-intent relevance loss in the training algorithm.
    \item \textbf{UO-ISC without text alignment and pretrained VLM:} This model utilizes UO-ISC but does not include the text alignment and the frozen pretrained CLIP modules. It retains the CLIP architecture but is trained with randomly initialized parameters.
    \item \textbf{Query-aware image semantic coding in~\cite{chen2024}}: This model transmits the answer generated by an LLM along with a compressed image to the receiver, and employs a pretrained diffusion model to reconstruct the image based on both the generated answer and the compressed image.
    \item \textbf{Traditional Image Coding (JPEG2000)~\cite{JPEG2000}:} As a conventional baseline, JPEG2000 is used to compress and reconstruct images without considering semantic relevance or user-specific intent.
\end{itemize}

\begin{figure}[t]
    \centering
    
    \begin{subfigure}[b]{0.235\textwidth}
        \centering
        \includegraphics[width=\textwidth]{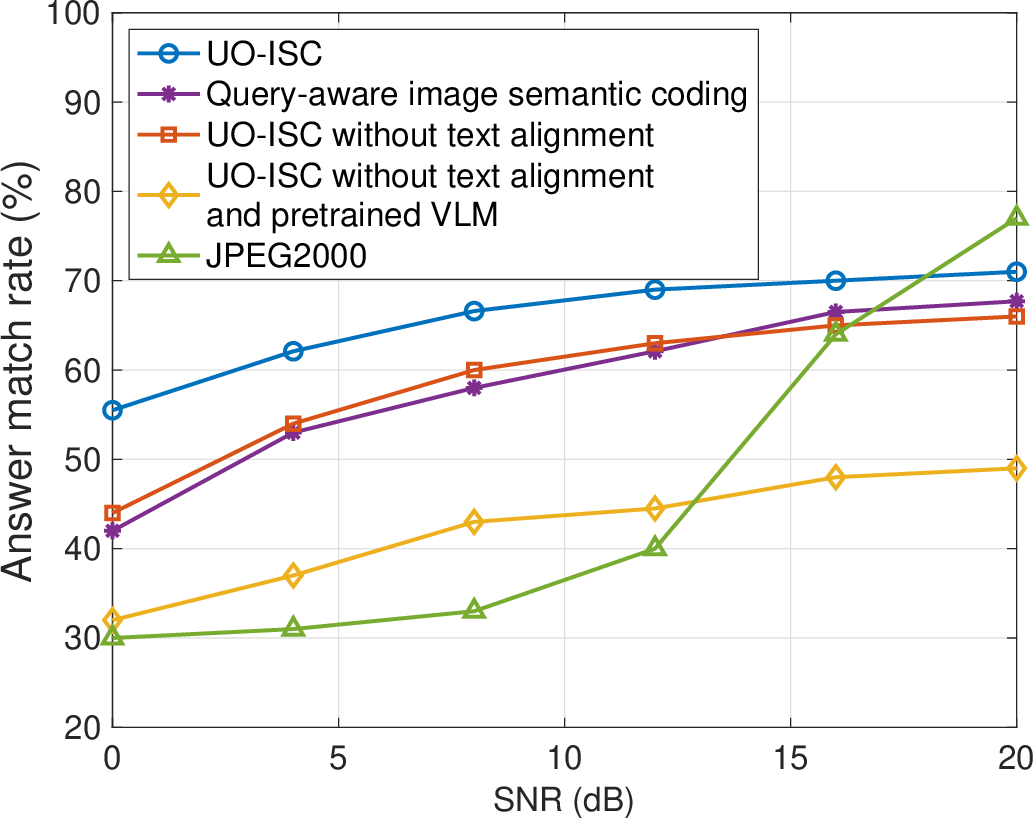}
        \caption{}
        \label{fig:suba}
    \end{subfigure}
    \hfill
    \begin{subfigure}[b]{0.235\textwidth}
        \centering
        \includegraphics[width=\textwidth]{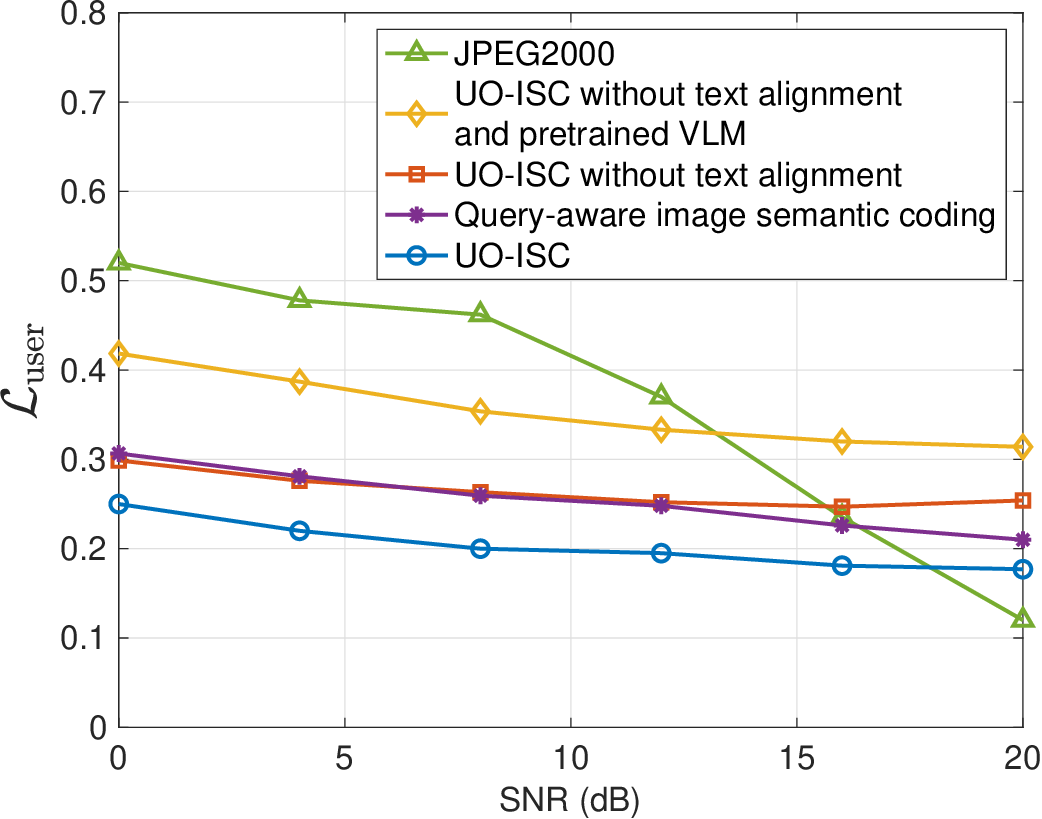}
        \caption{}
        \label{fig:subb}
    \end{subfigure}
    
    \caption{(a) Answer match rate and (b) user-intent relevance loss versus the SNR.}
    \label{fig:general}
    \vspace{-10pt}
\end{figure}

We evaluate both the generalization and text-alignment capabilities of the proposed UO-ISC framework. To do so, we train on a subset of the VQA dataset by excluding animal-related samples and perform zero-shot evaluation on a subset containing only animal-related samples. Fig.~\ref{fig:visual} shows the reconstructed images generated by UO-ISC and the baseline models with the signal-to-noise ratio (SNR) equals to 0 dB. Compared with the baselines, UO-ISC better preserves semantic features related to the user's prompt, as it transmits features related to the text query. On the other hand, the query-aware image semantic coding in~\cite{chen2024} hallucinates details based on the generated answer, while our baseline model without text alignment fails to capture query-related details. The model without a pretrained VLM overfits the training distribution and fails to generalize to unseen animal dataset, while JPEG2000 exhibits severe visual artifacts under low SNR.
Results in Figs.~\ref{fig:general}~(a) and (b) show the answer match rate and the user-intent relevance loss $\mathcal{L}_{\text{user}}$ versus the SNR. The answer match rate measures the proportion of samples where the answer generated by LLaVA from the reconstructed image exactly matches the one from the original image under the same user query. When the SNR is equal to 20 dB, UO-ISC achieves the lowest user-intent relevance loss and improves the answer match rate by 5\% through text alignment and by 34\% by using a large VLM backbone. It also outperforms the query-aware semantic coding baseline by 4.8\%.

\begin{figure}
    \centering
    \includegraphics[width=0.70\linewidth]{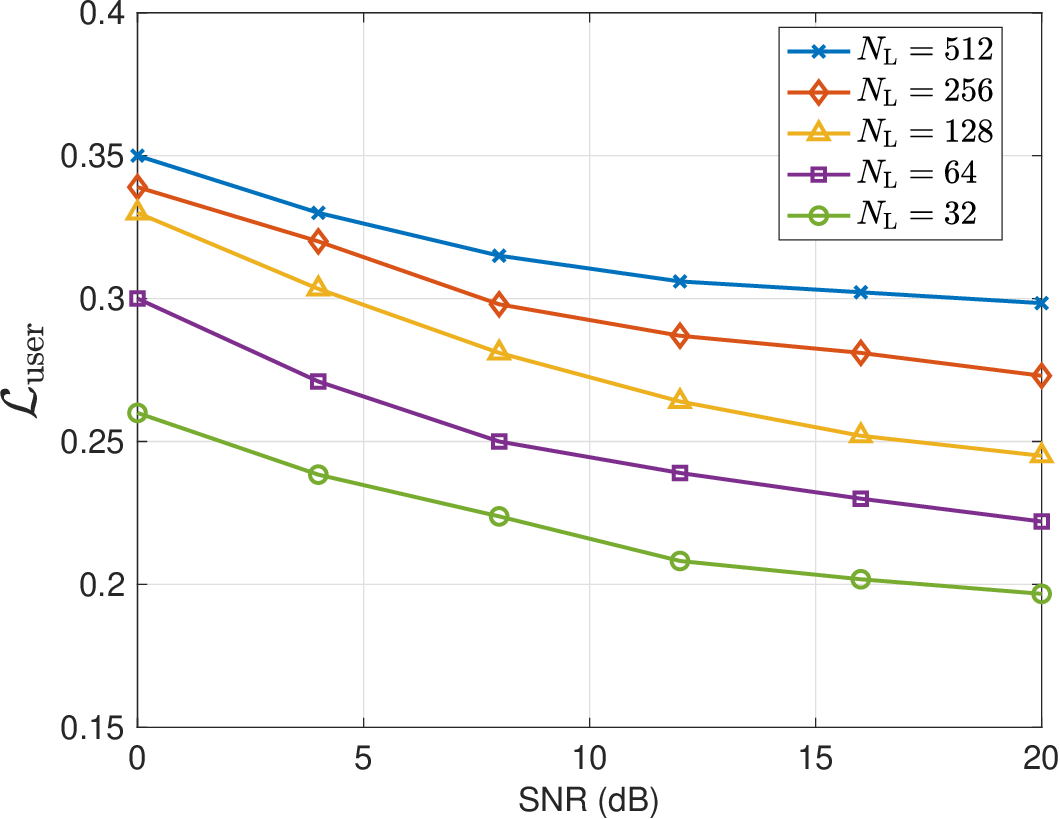}
    \caption{The user-intent relevance loss versus the SNR under different feature vector segment length $N_{\text{L}}$.}
    \label{fig:quant}
    \vspace{-20pt}
\end{figure}

\begin{figure}[t]
    \centering
    \includegraphics[width=0.70\linewidth]{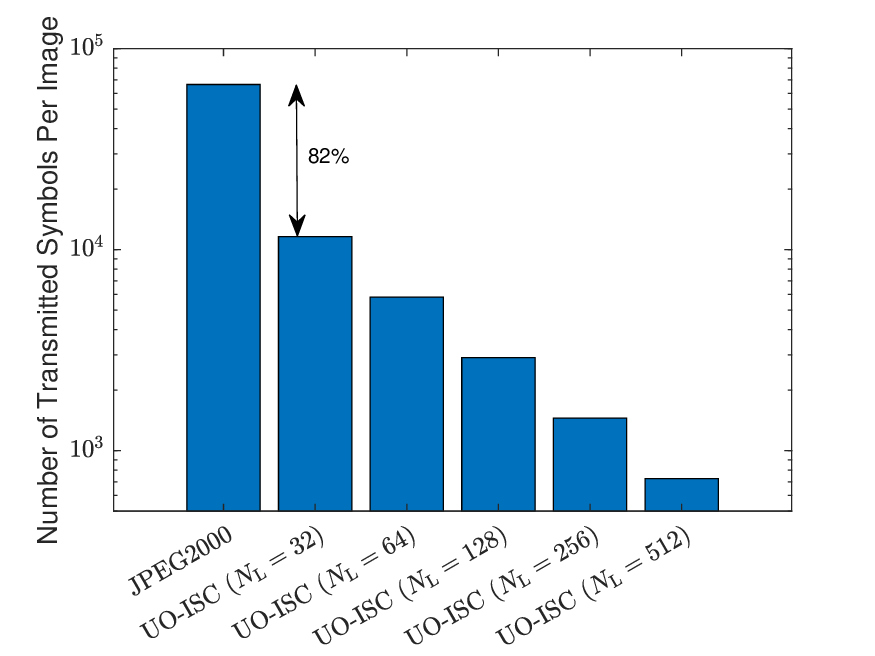}
    \caption{The number of transmitted symbols per image under different feature vector segment length $N_{\text{L}}$.}
    \label{fig:symbol}
    \vspace{-20pt}
\end{figure}

Fig.~\ref{fig:quant} illustrates the impact of the feature vector segment length $N_{\text{L}}$ on the user-intent relevance loss $\mathcal{L}_{\text{user}}$ under different SNR levels. A shorter segment length divides the feature vector into smaller segments, resulting in lower user-intent relevance loss $\mathcal{L}_{\text{user}}$ and more robust performance, since changes in a single segment affect only part of the feature vector. Although shorter segments can reduce the user-intent relevance loss $\mathcal{L}_{\text{user}}$, they increase the number of transmitted symbols. Fig.~\ref{fig:symbol} shows the number of transmitted symbols per image for various segment length. The number of transmitted symbols of our proposed UO-ISC framework is much lower than JPEG2000. When $N_{\text{L}}$ is equal to 32, the number of transmitted symbols of UO-ISC is 82\% lower than JPEG2000. This shows the efficiency of our proposed framework.


\section{Conclusion}
\label{Sec:conclude}
In this paper, we proposed a generalized UO-ISC framework. To improve generalization, we leveraged a pretrained large VLM in our framework to generate image and text features. To incorporate the user's intent, we aligned the image features with text features before image generation. Moreover, we introduced a segment-wise codebook for feature quantization and defined a user-intent relevance loss using a pretrained LLaVA model. Results on zero-shot experiments showed that using a pretrained VLM improves the generalization ability for unseen objects, while aligning text in image semantic coding enhances user-intent alignment. For future work, we will study how different types of user queries may influence the user-intent preservation.
\vspace{-2pt}

\bibliographystyle{ieeetr}
\bibliography{refs}
\end{document}